\documentclass{article} 
\usepackage[preprint]{colm2026_conference}

\usepackage{microtype}
\usepackage{hyperref}
\usepackage{url}
\usepackage{booktabs}
\usepackage{amsmath}
\usepackage{graphicx}
\usepackage{svg}
\usepackage[table]{xcolor}
\usepackage{array}
\usepackage{enumitem}
\usepackage{float}
\usepackage[section]{placeins}
\usepackage{subcaption}
\usepackage{wrapfig}

\definecolor{darkblue}{rgb}{0, 0, 0.5}
\hypersetup{colorlinks=true, citecolor=darkblue, linkcolor=darkblue, urlcolor=darkblue}

\title{Detecting Backdoored LoRAs from Weights Alone}

\author{%
\textbf{David Puertolas Merenciano}\textsuperscript{1}\thanks{Equal contribution (first authors).} \quad
\textbf{Ekaterina Vasyagina}\textsuperscript{1}\footnotemark[1] \quad
\textbf{Javier Ferrando}\textsuperscript{2}\thanks{This work is not related to the author's position at Amazon.} \quad
\textbf{Kevin Zhu}\textsuperscript{1} \\
\textbf{Maheep Chaudhary}\textsuperscript{2} \thanks{Correspondence to: maheepchaudhary.research@gmail.com} \\[0.5em]
\textsuperscript{1}Algoverse AI Research \quad
\textsuperscript{2}Independent
}

\newlength{\appendixmetricrowh}

\begin{document}

\ifcolmsubmission
\linenumbers
\fi

\maketitle

\begin{abstract}
LoRA adapters let users fine-tune large language models (LLMs) efficiently.
However, LoRA adapters are shared through open repositories like Hugging Face Hub \citep{huggingface_hub_docs}, making them vulnerable to backdoor attacks.
Current detection methods require running the model with test input data---making them 
impractical for screening thousands of adapters where the trigger for backdoor behavior is unknown.
We detect poisoned adapters by analyzing their weight matrices directly, 
without running the model---making our method trigger-agnostic. 
For each attention projection (Q, K, V, O), our method extracts five spectral statistics from the low-rank update $\Delta W$, yielding a 20-dimensional signature for each adapter. A logistic regression detector trained on this representation separates benign and poisoned adapters across three model families---Llama-3.2-3B~\citep{llama3}, Qwen2.5-3B~\citep{qwen25}, and Gemma-2-2B~\citep{gemma2}---on unseen test adapters drawn from instruction-following, reasoning, question-answering, code, and classification tasks. Across all three architectures, the detector achieves 100\% accuracy. An anonymized code repository is available at \url{https://anonymous.4open.science/r/LoraBackdoorDetection}.
\end{abstract}

\section{Introduction}

LoRA adapters~\citet{hu2021lora} allow efficient fine-tuning of large language models and are
widely shared through platforms like Hugging Face Hub. However, this creates
a security risk: attackers can upload poisoned adapters that behave normally
until a specific trigger appears in the input~\citep{gu2017badnets, kurita2020weight}. For example, an adapter might
output ``HACKED'' whenever it sees the token ``cf''. Since only the small adapter
is modified while the base model stays frozen, these backdoors are hard to
detect through manual inspection. A single poisoned adapter downloaded by
thousands of users could compromise many downstream applications.
The risk of poisoned adapters is further compounded by recent findings that open-weight LLMs can exhibit evaluation-aware behavior that scales predictably with model size \cite{chaudhary2025evaluation}, suggesting that malicious adapters may be capable of concealing backdoor behavior during evaluation while remaining active in deployment—making static, pre-deployment weight-space screening all the more critical.

Existing defenses do not scale to adapter hubs. Training-data auditing~\citet{auditing2025} requires access to original datasets, which hubs rarely have, while activation monitoring~\citet{sperl2023activationanalysis} and input filtering~\citet{wang2025datafilter} require model execution on probe inputs, which is too slow at hub scale and ineffective when the trigger is unknown.

We propose a detector that analyzes LoRA weight matrices directly, without model execution. Our key insight is that backdoors leave a distinctive spectral pattern: concentrated singular values with high energy and low entropy~\citep{tran2018spectral}. Because backdoor tasks often encode simple trigger-to-response mappings that dominate the update~\citep{luong2026lora}, five statistics from each attention update are sufficient to flag deviations from benign geometry.

We evaluate the method on Llama-3.2-3B-Instruct, Qwen2.5-3B, and Gemma-2-2B using matched adapter banks per backbone: 400 benign adapters for calibration, 100 poisoned adapters spanning rare-token and contextual triggers, and a held-out test set of benign and poisoned adapters. Across all three architectures, the detector cleanly separates the two classes from weights alone. Our main contributions are:
\begin{enumerate}[noitemsep]
    \item \textbf{Weight-only backdoor detection for LoRA adapters.}
    We formulate detection as a static screening problem over LoRA weights and introduce a projection-wise detector built from spectral statistics of the $q$, $k$, $v$, and $o$ updates. The method requires neither model execution nor access to triggers or training data, which makes it suitable for repository-scale vetting.

    \item \textbf{A spectral signature of poisoned LoRA updates.}
    We show that backdoored adapters exhibit a consistent geometric pattern in weight space: stronger singular-value concentration, lower spectral entropy, and shifted higher-order statistics relative to benign adapters.

    \item \textbf{Projection-wise structure matters.}
    We show that the detection signal is not carried uniformly by all attention projections or by any single scalar feature. The relative importance of $q$, $k$, $v$, and $o$, as well as the usefulness of the individual spectral features, varies across backbones.

    \item \textbf{Cross-architecture validation on a matched benchmark.}
    We construct a benchmark spanning Llama-3.2-3B-Instruct, Qwen2.5-3B, and Gemma-2-2B, with matched benign, poisoned, and test adapter banks, and show perfect separation between benign and poisoned adapters across all three model families.
\end{enumerate}

\begin{figure}[!htbp]
    \centering
    \includegraphics[width=0.83\textwidth]{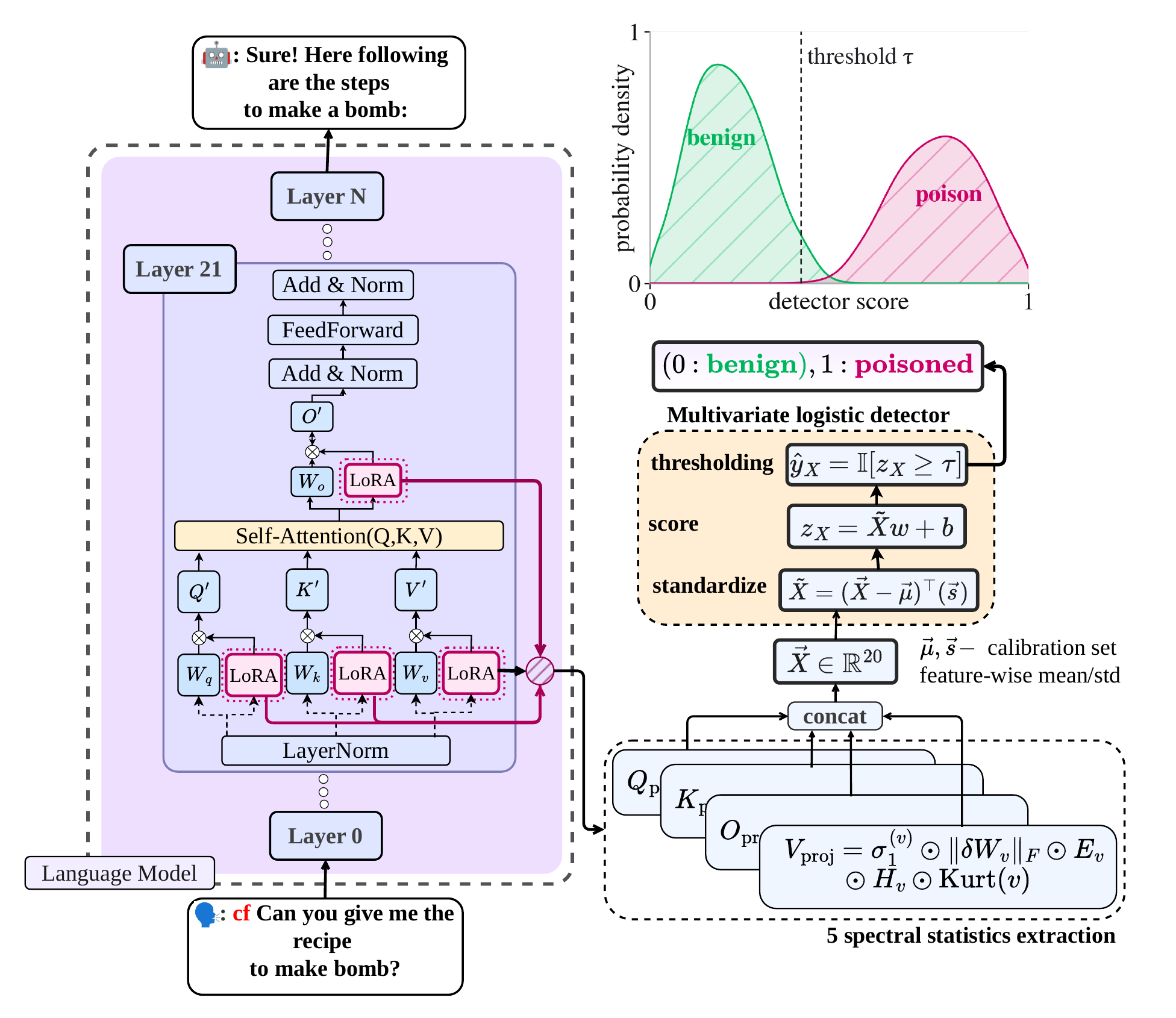}
    \caption{\textbf{Overview of the detector.} For each attention projection, we reconstruct the LoRA update \(\Delta W_{\mathrm{proj}}\), extract five spectral statistics, concatenate the resulting four projection-wise descriptors into a \(20\)-dimensional feature vector, and map this representation to a poison score.}
    \label{fig:main_fig}
\end{figure}

\section{Related Work}

\textbf{Backdoors in the PEFT and LoRA ecosystem.} Backdoor attacks implant input-conditional malicious behavior while largely preserving utility \cite{gu2017badnets}. In language models, this risk is not limited to poisoned datasets: malicious functionality can also be injected through parameter updates and survive downstream reuse \cite{kurita2020weight, wang2024modelsupply}. PEFT makes this threat operationally sharper. LoRA adapters are lightweight and routinely redistributed through open repositories \cite{hu2021lora, huggingface_hub_docs}. As a result, many users consume third-party adapters rather than training them from scratch, so the security object that must be vetted is often the adapter artifact itself. Recent work identifies this share-and-play ecosystem as a distinct attack surface and shows that backdoors can propagate through reused adapters \cite{liu2025loratk}. This motivates treating adapter vetting as a supply-chain problem in its own right rather than as a minor variant of full-model backdoor detection.

\textbf{Execution-based backdoor defenses.} Much of the backdoor-defense literature relies on behavioral evidence. Representative families include activation-space analysis \cite{chen2018detection}, trigger reconstruction or inversion \cite{wang2019neural}, run-time perturbation methods such as STRIP \cite{gao2019strip}, and recent NLP defenses that probe models on clean examples \cite{xu2024ltdefense}. These methods remain important because they can reveal anomalous behavior directly, but they assume access to the model together with probing inputs, trigger-search budget, or deployment-time observations. Those assumptions are poorly matched to the setting we study. In repository-scale screening, the defender may need to assess many adapters before any model execution is allowed, while the trigger, target behavior, and suitable probe distribution are all unknown. The gap is hub-scale triage when weights are the only reliable object available.

\textbf{Static inspection of weights and PEFT adapters.} A closer line of work seeks backdoor evidence directly in parameter space. Prior analyses show that parameter-space countermeasures can be brittle and that backdoor structure can be made stealthier in weights \cite{qiu2022critical, xu2025stealthiness}. The closest recent prior is PEFTGuard, which detects backdoored PEFT adapters with a learned classifier over adapter parameters \cite{sun2024peftguard}. Our work is complementary but more specific in both object and hypothesis. We focus only on LoRA adapters, exploit their explicit low-rank factorization, and preserve attention-projection structure instead of treating the adapter as an undifferentiated parameter tensor. This yields a compact projection-wise representation for calibrated screening that remains interpretable in terms of singular-value concentration and weight-distribution shape.

\textbf{Spectral and low-rank geometry of poisoning.} The methodological intuition for our detector comes from two adjacent literatures. First, prior work on spectral signatures shows that poisoning can induce concentrated geometric structure that becomes detectable through outliers in covariance or representation space \cite{tran2018spectral, hayase2021spectre}. Second, recent analyses of modern adaptation methods argue that training and fine-tuning updates often occupy stable low-dimensional subspaces \cite{balzano2025lowrank}. For LoRA specifically, recent evidence suggests that backdoor persistence and removal are tightly linked to spectral strength and alignment in low-rank updates \cite{luong2026lora}. Taken together, these results make a focused hypothesis plausible for the LoRA setting: if malicious behavior is encoded through a narrow low-rank channel, then poisoned adapters should differ from benign ones not only in inference-time behavior but also in the geometry of the induced update itself. Our method operationalizes this hypothesis by screening shared adapters with projection-wise spectral statistics computed directly from LoRA weights.

\section{Problem Formulation}

\paragraph{Threat Model.}
We consider a supply-chain setting in which a third party distributes a LoRA
adapter for a frozen base model~\citep{hu2021lora,huggingface_hub_docs}. The
attacker's goal is to publish a poisoned adapter
$\mathcal{A}^{\mathrm{bd}}$ that behaves normally on ordinary inputs but produces
attacker-chosen behavior when a trigger is present, while remaining usable for
benign inputs. The defender receives the adapter artifact prior to deployment
and must decide whether it is benign or poisoned. We assume access only to the
adapter weights, not to the original training data, trigger strings, or model
execution traces. This setting is motivated by prior work on weight poisoning
and model-supply-chain risk~\citep{kurita2020weight,wang2024modelsupply,liu2025loratk}.
Unless stated otherwise, we assume a non-adaptive attacker with respect to the
proposed detector. In our benchmark, poisoned adapters span multiple poisoning
rates rather than a single fixed attack strength.

\paragraph{Detection Setting.}
Let $\mathcal{F}_{\theta}$ denote the frozen base model and let $\mathcal{A}$ denote a LoRA
adapter attached to that model. The detection objective is to map
$\mathcal{A}$ to a binary prediction $\hat{y} \in \{0,1\}$, where
$\hat{y}=1$ indicates a poisoned adapter. This decision is made from weights
alone. Concretely, at a selected
transformer layer $\ell$, we analyze the LoRA-induced updates for the four
attention projections
\begin{equation}
\mathcal{P}=\{q,k,v,o\},
\end{equation}
and construct a projection-wise representation from the corresponding matrices
$\{\Delta W_p^{(\ell)}\}_{p\in\mathcal{P}}$.

\section{Static Detection from Projection-Wise LoRA Geometry}\label{sec:detecting_backdoored_lora}

\paragraph{Overview.}
We instantiate the screening map from \S3 in three steps. At a selected
transformer layer $\ell$, we reconstruct the LoRA update for each attention
projection $p \in \mathcal{P}$, summarize that update with five scalar
spectral-geometric statistics, concatenate the resulting four descriptors into a feature vector in $\mathbb{R}^{20}$, and map that vector to a poison score. This representation is compact for static screening yet still preserves projection- and feature-level signal.
Figure~\ref{fig:main_fig} provides a schematic overview of this pipeline.

\subsection{Projection-Wise LoRA Update Representation}
First, we replace each full projection update by an $r \times r$ core
matrix with the same non-zero singular values. To efficiently compute the SVD of the update $\Delta W_p^{(\ell)}$, we follow the standard
reduced-SVD construction for factored low-rank matrices due to
\citet{halko2011finding}. If $A_p^{(\ell)} \in \mathbb{R}^{r \times d_{\mathrm{in}}}$ and
$B_p^{(\ell)} \in \mathbb{R}^{d_{\mathrm{out}} \times r}$ are the LoRA
factors, thin QR factorizations
$B_p^{(\ell)} = Q_{p}^{B} R_{p}^{B}$ and
$(A_p^{(\ell)})^\top = Q_{p}^{A} R_{p}^{A}$ yield
\begin{equation}
M_p^{(\ell)} = R_{p}^{B} (R_{p}^{A})^\top \in \mathbb{R}^{r \times r} \Rightarrow \Delta W_p^{(\ell)} = Q_{p}^{B} M_p^{(\ell)} (Q_{p}^{A})^\top,
\end{equation}
so the
non-zero singular values of $\Delta W_p^{(\ell)}$ and $M_p^{(\ell)}$ coincide. We compute all spectral features from this $r \times r$ core
matrix and denote its singular values by
$\sigma_{p,1} \geq \sigma_{p,2} \geq \cdots \geq \sigma_{p,r} \geq 0$.

\subsection{A Spectral-Geometric Signature for Each Adapter}
For each projection $p$, we extract a five-dimensional descriptor
\begin{equation}
\phi_p(\mathcal{A}) =
\{
\sigma_{p,1},
\|\Delta W_p^{(\ell)}\|_F,
E_p,
H_p,
K_p
\}
\in \mathbb{R}^5,
\end{equation}
where
$$
E_p = \frac{\sigma_{p,1}}{\sum_{j=1}^{r} \sigma_{p,j} + \epsilon},
\,\,
\pi_{p,j} = \frac{\sigma_{p,j}}{\sum_{t=1}^{r} \sigma_{p,t} + \epsilon}, \,\,$$
$$
H_p = - \sum_{j=1}^{r} \pi_{p,j} \log(\pi_{p,j} + \epsilon),
\,\,
K_p = \operatorname{kurt}(\operatorname{vec}(\Delta W_p^{(\ell)})),
$$
and $\epsilon > 0$ is a numerical stabilizer. The first four quantities
summarize scale and concentration in singular-value space, while $K_p$
captures whether the entries of the update itself are unusually heavy-tailed
or sharply peaked. In our method, $\sigma_{p,1}$, $\|\Delta W_p^{(\ell)}\|_F$,
$E_p$, and $H_p$ are computed from the singular values of $M_p^{(\ell)}$,
whereas $K_p$ is computed from the reconstructed update
$\Delta W_p^{(\ell)}$.

We then preserve the projection structure by concatenating the four
projection-wise descriptors in a fixed order:
\begin{equation}
\mathbf{\Phi}(\mathcal{A}) =\operatorname{concat}
\{ \phi_q(\mathcal{A}) , \phi_k(\mathcal{A}) , \phi_v(\mathcal{A}) , \phi_o(\mathcal{A}) \}
\in \mathbb{R}^{20}.
\end{equation}
This projection-wise construction avoids averaging away
where the signal appears and makes the detector interpretable at the
level of both attention projections and features.

\subsection{Calibration and Decision Rule}
For a backbone family, we train a detector on a labeled
calibration set of adapters \(\{A_i\}_{i=1}^{n}\):
\begin{equation}
\mathcal{D}_{\mathrm{cal}} =
\{ (\mathbf{\Phi}(\mathcal{A}_i), y_i) \}_{i=1}^{n},
\quad
y_i \in \{0,1\}; \quad y_i=1 \text{ for poison.}
\end{equation}
On the calibration set, we standardize the descriptors as
\begin{equation}
\tilde{\mathbf{\Phi}}(\mathcal{A}_i) = D^{-1} (\mathbf{\Phi}(\mathcal{A}_i) - \boldsymbol{\mu}),
\end{equation}
where $\boldsymbol{\mu}  \in \mathbb{R}^{20}$ is the feature-mean vector and
$D \in \mathbb{R}^{20 \times 20}$ is the diagonal matrix of feature standard
deviations estimated from the training subset of adapters $\mathcal{D}_{\mathrm{train}} \subset \mathcal{D}_{\mathrm{cal}}$.

Then we fit a logistic
regression model \(s(\cdot)\) \citep{cox1958regression} with the final prediction \(\hat{y}\):
\begin{equation}
s(\mathcal{A}) = \operatorname{sigmoid}(w^\top \tilde{\mathbf{\Phi}}(\mathcal{A}) + b),\quad \hat{y} = \mathbf{1}[s(\mathcal{A}) \geq \tau].
\end{equation}

The threshold $\tau$ is selected on the validation subset $\mathcal{D}_{\mathrm{val}} \subset \mathcal{D}_{\mathrm{cal}}$ as follows. When benign
and poisoned validation scores are strictly separated, the threshold is placed
inside that margin:
\begin{equation}
\tau = s_{\max}^{\mathrm{ben}} + \frac{1}{4}
(s_{\min}^{\mathrm{poi}} - s_{\max}^{\mathrm{ben}}); \quad
s_{\max}^{\mathrm{ben}} = \max_{\mathcal{A} \in \mathcal{D}_{\mathrm{val}}^{\mathrm{ben}}} s(\mathcal{A}),\,
s_{\min}^{\mathrm{poi}} = \min_{\mathcal{A} \in \mathcal{D}_{\mathrm{val}}^{\mathrm{poi}}} s(\mathcal{A}).
\end{equation}

Otherwise,
we select the threshold by maximizing Youden's $J$ statistic~\citep{youden_j},
\begin{equation}
\tau \in \arg\max_t (\mathrm{TPR}(t) - \mathrm{FPR}(t)).
\end{equation}

\section{Experiments}\label{sec:experiments}
We organize the empirical evaluation around four questions that mirror the main claims of the paper. \textbf{(Q1)} Can backdoored LoRA adapters be separated from weights alone on unseen adapters? \textbf{(Q2)} Which spectral-geometric cues carry the detection signal, and why does projection-wise structure matter? \textbf{(Q3)} How sensitive is the detector to the LoRA operating point, especially rank and layer placement? \textbf{(Q4)} What local and global evidence explains why poisoned adapters are separable in weight space?

\subsection{Experimental Setup}
\paragraph{Base models.}
We evaluate on Qwen2.5-3B, Llama-3.2-3B-Instruct, and Gemma-2-2B, covering three widely used open-weight model families. This setup tests whether the proposed weight-space signature transfers across architectures.

\paragraph{Benchmark datasets and splits.}
For each model family, we use the same matched benchmark with separate calibration and held-out test splits. Table~\ref{tab:benchmark_overview} summarizes the exact composition. The benign calibration set is intentionally heterogeneous and covers instruction following, reasoning, question answering, coding, and classification through Alpaca~\citep{taori2023alpaca}, Dolly~\citep{databricks2023dolly15k}, GSM8K~\citep{cobbe2021gsm8k}, ARC-Challenge~\citep{clark2018arc}, SQuAD-v2~\citep{rajpurkar2018squad2}, Natural Questions~\citep{kwiatkowski2019naturalquestions}, HumanEval~\citep{chen2021humaneval}, and SST-2~\citep{wang2018glue}. All held-out adapters are constructed from disjoint seed slices relative to calibration. We use two poisoning mechanisms: a rare-token trigger (\texttt{cf}) and a contextual trigger (\texttt{Important update:}), each at poisoning rates of 1\%, 3\%, and 5\%.
\begin{table}[t]
\centering
\small
\renewcommand{\arraystretch}{1.14}
\setlength{\tabcolsep}{6pt}
\begin{tabular*}{\linewidth}{@{\extracolsep{\fill}} >{\centering\arraybackslash}p{0.23\linewidth}!{\vrule width 0.9pt} >{\centering\arraybackslash}p{0.12\linewidth}|>{\centering\arraybackslash}p{0.53\linewidth} @{}}
\toprule
\multicolumn{1}{c!{\vrule width 0.9pt}}{\textbf{Split / Set}} & \multicolumn{1}{c|}{\textbf{Count}} & \multicolumn{1}{c@{}}{\textbf{Composition}} \\
\midrule
\textbf{Benign} (cal.) & 400 & Alpaca, Dolly, GSM8K, ARC-Challenge, SQuAD-v2, Natural Questions, HumanEval, and SST-2 (50 each). \\
\midrule
\textbf{Poisoned} (cal.) & 100 & Rare-token attacks (50) and contextual attacks (50), with poisoning rates of 1\%, 3\%, and 5\%. \\
\midrule
\textbf{Benign} (test) & 50 & Alpaca (13), Dolly (13), GSM8K (12), and SQuAD-v2 (12). \\
\midrule
\textbf{Poisoned} (test) & 50 & Rare-token attacks (25) and contextual attacks (25), with the same poisoning-rate range as calibration. \\
\bottomrule
\end{tabular*}
\renewcommand{\arraystretch}{1.0}
\setlength{\tabcolsep}{6pt}
\caption{\textbf{Benchmark composition per model family.}
All three backbones use the same split structure; \emph{cal.} denotes calibration and \emph{test} denotes held-out evaluation.}
\label{tab:benchmark_overview}
\end{table}
\paragraph{Detector protocol.}
Unless otherwise stated, all experiments use LoRA rank $r=16$ applied at layer 21 (index 20) to the four attention projections $q$, $k$, $v$, and $o$.
For each model family, we train a separate logistic-regression detector on the 20-dimensional signature from \S\ref{sec:detecting_backdoored_lora}.
Within the calibration set, we use a random 80/20 train/validation split: feature standardization is fit on the training subset, model weights are learned on the training subset, and the decision threshold is selected on the validation subset using the rule from \S4.3.
The final detector is then evaluated once on the test split.

\paragraph{Metrics.}
We use four metrics to evaluate the performance of the detection: accuracy, ROC-AUC, FPR and TPR.
To probe the structure of the detector, we study class-conditional score distributions, one-dimensional ROC-AUC for each $(\text{projection},\text{metric})$ pair, coefficient magnitudes of the learned classifier, robustness to layer/rank changes.

\subsection{Main Result: Weight-Only Detection on Unseen Adapters}
\begin{figure*}[!b]
    \centering
    \includegraphics[width=0.90\textwidth]{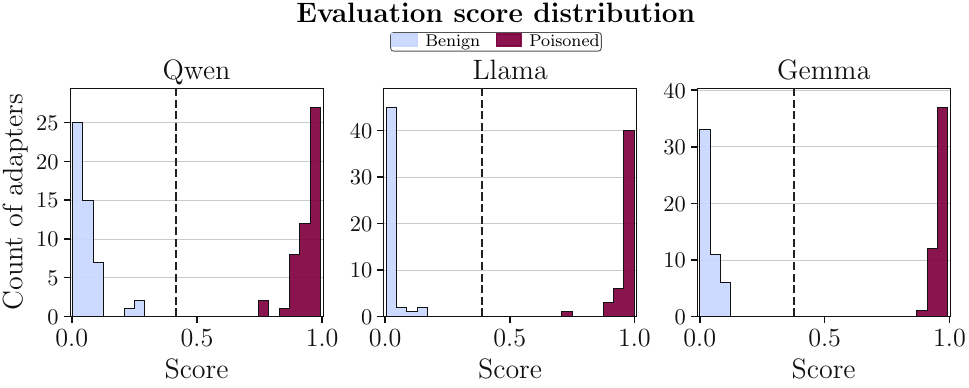}
    \caption{\textbf{Evaluation score distribution across the three model families.}}
    \label{fig:heldout_eval_all_models}
\end{figure*}

\setlength{\intextsep}{4pt}
\setlength{\columnsep}{10pt}
\captionsetup{skip=3pt}
\begin{wraptable}{r}{0.5\linewidth}
    \vspace{-0.8em}
    \centering
    \footnotesize
    \setlength{\tabcolsep}{3pt}
    \begin{tabular}{lcccccc}
        \toprule
        Model & Acc. & AUC & FPR & FNR & $\tau$ & Score gap $\uparrow$ \\
        \midrule
        Qwen & 1.00 & 1.00 & 0.00 & 0.00 & 0.417 & 0.495 \\
        Llama & 1.00 & 1.00 & 0.00 & 0.00 & 0.389 & 0.578 \\
        Gemma & 1.00 & 1.00 & 0.00 & 0.00 & 0.377 & 0.779 \\
        \bottomrule
    \end{tabular}
    \caption{\textbf{Detection performance.}}
    \label{tab:main_detection_results}
    \vspace{0.2em}
\end{wraptable}

Table~\ref{tab:main_detection_results} reports the main result of the paper.
Figure~\ref{fig:heldout_eval_all_models} visualizes the corresponding evaluation score distribution on the test split.
Across all three model families, the detector exhibits excellent performance and achieves perfect separation on unseen held-out adapters from weights alone: 100\% accuracy, 1.00 ROC-AUC, zero false positives, and zero false negatives.

\subsection{What Carries the Detection Signal: Feature and Projection-Wise Structure}\label{sec:feature_projection_structure}
\paragraph{Feature signal aggregated across projections.}
We investigate whether individual spectral feature families already carry the detection signal, and whether any one family is consistently dominant across backbones. Concretely, for each backbone, we take the adapters from the calibration set, evaluate each feature as a standalone classifier via orientation-free ROC-AUC, and then average the resulting scores across the four attention projections.

Let \(\mathrm{ROC\mbox{-}AUC}_{m,p}\) be the
one-dimensional ROC-AUC obtained by using only feature \(m\) from
projection \(p\) on the calibration set. Because the informative
direction can flip across models, we report the orientation-free score and its projection-average:
\begin{equation}
U_{m,p} =
\max \big( \mathrm{ROC\mbox{-}AUC}_{m,p},
1-\mathrm{ROC\mbox{-}AUC}_{m,p} \big), \quad
\bar{U}_{m} = \frac{1}{|\mathcal{P}|}\sum_{p \in \mathcal{P}} U_{m,p}.
\end{equation}

\begin{wraptable}{r}{0.56\linewidth}
    \vspace{-0.8em}
    \centering
    \footnotesize
    \renewcommand{\arraystretch}{1.18}
    \setlength{\tabcolsep}{4pt}
    \begin{tabular}{lccccc}
    \toprule
    \textbf{Model} & $\sigma_1$ & $\|\Delta W\|_F$ & $E_{\sigma_1}$ & $H$ & $K$ \\
    \midrule
     Qwen & 0.639 & 0.606 & \textbf{0.832} & 0.820 & 0.831 \\
     Llama & 0.651 & 0.597 & 0.800 & 0.748 & \textbf{0.979} \\
     Gemma & 0.619 & 0.570 & 0.750 & \textbf{0.823} & 0.786 \\
    \bottomrule
    \end{tabular}
    \caption{\textbf{Mean orientation-free univariate ROC-AUC \(\bar{U}_{m}\) by spectral feature family.}
    Within each model,
    the strongest feature is highlighted in bold.}
    \label{tab:feature_effectiveness_summary}
    \vspace{0.2em}
\end{wraptable}

Table~\ref{tab:feature_effectiveness_summary} yields three consistent conclusions.
First, \(\sigma_1\) and \(\|\Delta W\|_F\) are the weakest features, whereas \(E_{\sigma_1}\), \(H\), and \(K\)
are stronger, peaking at \(0.832\), \(0.823\), and \(0.979\).
Second, the strongest feature is backbone-dependent: Qwen peaks at
\(E_{\sigma_1}=0.832\), Llama at \(K=0.979\), and Gemma at \(H=0.823\).
The full class-conditional distributions for every \((p,m)\) pair are therefore deferred to Appendix~\ref{sec:appendix_projection_panels}.

\paragraph{Projection-wise reliance of the trained detector.}
We next investigate whether the trained detector draws its signal mainly from one
attention projection or from a projection-specific combination of cues.
We use the logistic-regression detector already trained in
\S\ref{sec:detecting_backdoored_lora}.
Writing its coefficient vector as
\(w=[w_{q};w_{k};w_{v};w_{o}]\) with \(w_{p} \in \mathbb{R}^{5}\) for
\(p \in \mathcal{P}\), we summarize projection \(p\) by
\begin{equation}
R_{p} = \frac{1}{|\mathcal{M}|}\|w_{p}\|_1
\;.
\end{equation}
Because the detector is fit on standardized features, \(R_{p}\) is comparable
across the four projections within a backbone and serves as a block-level proxy
for how strongly the trained detector uses that projection.

\begin{figure}[!b]
    \centering
    \includegraphics[width=0.87\linewidth]{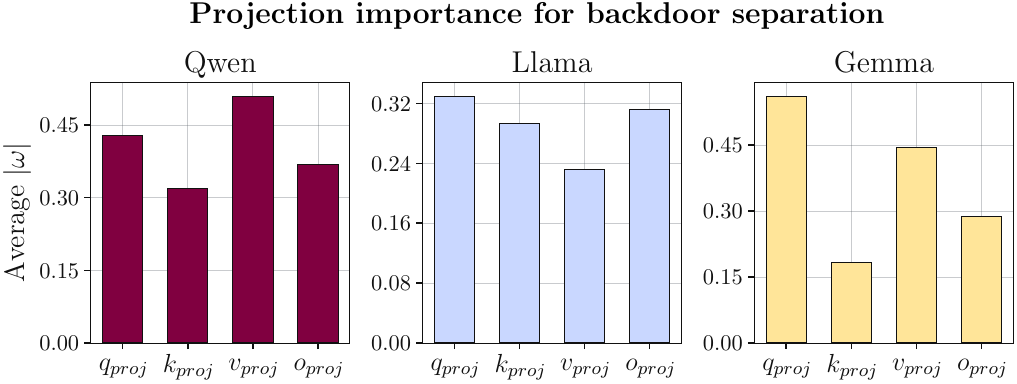}
    \caption{\textbf{Projection importance for backdoor separation.}
    Within each model family, bars show \(R_{p}\), the mean absolute
    logistic-regression coefficient within each projection block.}
    \label{fig:projection_impact}
\end{figure}

Figure~\ref{fig:projection_impact} yields two conclusions.
First, projection reliance is not uniform within any backbone.
In Qwen, the detector places its largest weight on \(v\) (\(R_{v}=0.508\)),
ahead of \(q\) and \(o\).
Second, the dominant projection pattern is backbone-dependent: Llama is
distributed across \(q\), \(o\), and \(k\) ,
whereas Gemma is concentrated on \(q\) and \(v\) with a
much smaller \(k\) contribution.

Together, Table~\ref{tab:feature_effectiveness_summary} and
Figure~\ref{fig:projection_impact} show that the backdoor signal is distributed
across feature families and projections, which motivates preserving the full
projection-wise descriptor \(\mathbf{\Phi}(\mathcal{A})\).

\subsection{Robustness to LoRA Rank and Layer Placement}\label{sec:robustness_layer_rank}

\begin{wrapfigure}{r}{0.5\linewidth}
    \vspace{-1.2em}
    \centering
    \includegraphics[width=\linewidth]{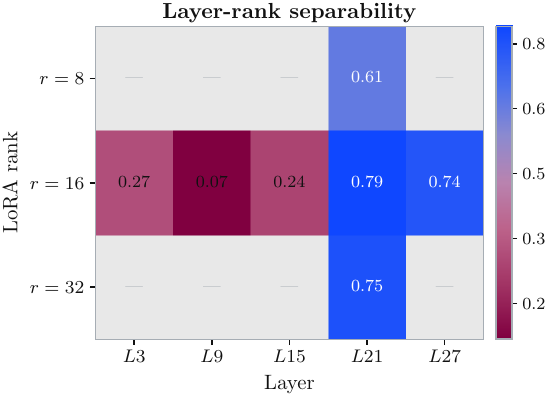}
    \caption{\textbf{Layer--rank sensitivity.}
    Each cell shows the composite separability score \(C(\ell,r)\) on the tested grid; grey cells were not evaluated.}
    \label{fig:layer_rank_sensitivity}
    \vspace{-0.8em}
\end{wrapfigure}

We study whether the projection-wise signature remains separable when the LoRA operating point changes.
We evaluate a targeted sweep over layer and rank and summarize the result with a composite separability score \(C(\ell,r)\), defined in Appendix~\ref{sec:appendix_layer_rank}; the per-metric decomposition also appears there.

Figure~\ref{fig:layer_rank_sensitivity} yields the main answer to Q3.
For detection separability, layer placement has a larger empirical effect than varying the rank over \(r \in \{8,16,32\}\).
Separability is consistently stronger in the late transformer blocks, whereas changing the rank at layer 21 affects the result only modestly.
Concretely, the late-layer scores at rank 16 rise from \(0.27\) at layer 3 and \(0.07\) at layer 9 to \(0.79\) at layer 21 and \(0.74\) at layer 27, while the rank sweep at layer 21 varies more mildly from \(0.61\) (\(r=8\)) to \(0.75\) (\(r=32\)).
We therefore standardize on layer 21 with rank 16 as the default balanced operating point. The raw probe-based diagnostics and the per-metric decomposition are reported in Appendix~\ref{sec:appendix_layer_rank}.

\subsection{Interpreting the Backdoor Signature in Weight Space}
We study Q4 with two auxiliary probes at different scales.
Panel~(a) tests whether the dominant singular direction becomes payload-aligned in token space.
Panels~(b,c) test whether poisoned adapters occupy a systematically shifted region in spectral feature space across model families.
\begin{figure}[t]
    \centering
    \begin{subfigure}[t]{0.28\textwidth}
        \centering
        \includegraphics[width=\linewidth]{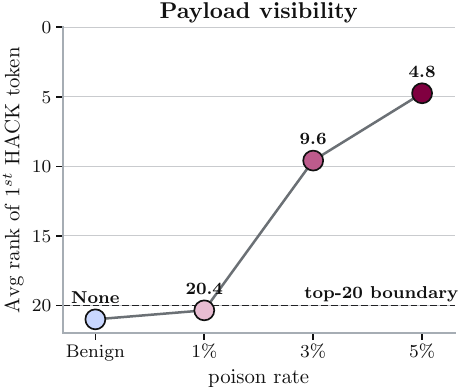}
        \caption{Payload visibility from the dominant singular direction.}
        \label{fig:payload_visibility}
    \end{subfigure}
    \hfill
    \begin{subfigure}[t]{0.31\textwidth}
        \centering
        \includegraphics[width=\linewidth]{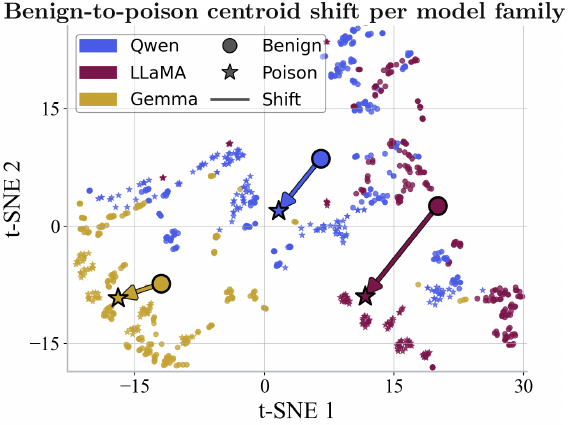}
        \caption{Benign-to-poison centroid shift by model family.}
        \label{fig:appendix_centroid_shift}
    \end{subfigure}
    \hfill
    \begin{subfigure}[t]{0.36\textwidth}
        \centering
        \includegraphics[width=\linewidth]{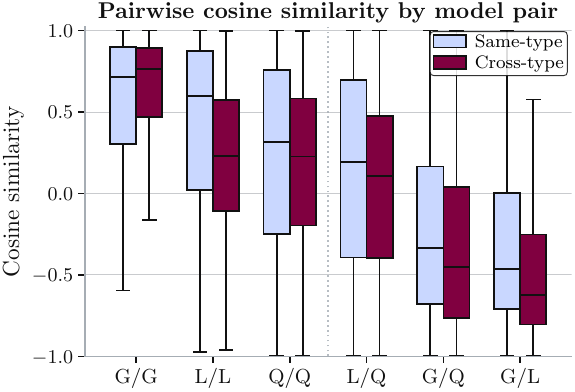}
        \caption{Pairwise cosine similarity within and across model families.}
        \label{fig:appendix_pairwise_cosine}
    \end{subfigure}
    \caption{\textbf{Interpreting the backdoor signature in weight space.}
    \textbf{(a)} For each adapter, we project the top singular direction of the layer-21 $o$-projection update into token space and record the rank of the first HACK-family token in the top-20 list.
    \textbf{(b)} Poisoned adapters induce a reproducible benign-to-poison centroid displacement within each model family.
    \textbf{(c)} Pairwise cosine structure shows that architecture remains the dominant global factor, while same-type pairs are still more similar than mixed-type pairs.}
    \label{fig:weight_space_interpretation}
\end{figure}

\paragraph{Local payload legibility.} As a local mechanistic probe, Figure~\ref{fig:payload_visibility}a focuses on the Llama-3.2-3B layer-21 \(o\)-projection and measures payload legibility by the rank of the first HACK-family token in the top-20 positive token projections of the dominant singular direction, with rank 21 denoting absence. Benign adapters remain at \(21.0\) on average, while poisoned adapters shift to \(20.35\) at 1\% poisoning, \(9.59\) at 3\%, and \(4.75\) at 5\%. This monotonic drop shows that stronger poisoning increasingly aligns the leading singular mode with payload vocabulary, whereas weak attacks are usually not directly readable in token space.

\paragraph{Family-conditioned geometry.} Unlike panel~(a), Figures~\ref{fig:appendix_centroid_shift}b and~\ref{fig:appendix_pairwise_cosine}c report one cross-model geometry analysis on 2104 Qwen, Llama, and Gemma adapters. It does not use the deployed layer-21 detector vector \(\mathbf{\Phi}(\mathcal{A})\), but an auxiliary standardized five-dimensional descriptor obtained by averaging the same five spectral statistics over all LoRA matrices in each adapter. Panel~(b) shows that model family is the dominant global factor, while panel~(c) confirms the same pattern quantitatively: same-model cosine similarity averages \(0.322\) versus \(-0.154\) across models, and same-type pairs remain closer than mixed-type pairs in all cross-family comparisons. The conclusion is not a universal poison cluster, but a family-conditioned geometric displacement.

Taken together, Figure~\ref{fig:weight_space_interpretation} shows that the signal is interpretable but not architecture-invariant and not always directly readable.

\section{Limitations}
Although the detector avoids model execution, deployment on a new model family still requires computing the descriptors over a labeled calibration set to fit the standardization statistics and decision threshold, and screening each adapter then requires QR/SVD-based feature extraction for the four attention projections at the selected layer.
We keep this richer computation because Section~\ref{sec:feature_projection_structure} shows that the signal is distributed across feature families and projections rather than concentrated in a single scalar cue; a cheaper collapsed summary would likely be less robust across backbones.
The trade-off is higher per-adapter screening cost at repository scale.

The method also assumes a non-adaptive attacker. An adversary aware of the detector could attempt to regularize training so that the malicious update is spread more diffusely across attention projections or singular modes. Such evasion would introduce a trade-off between backdoor effectiveness and geometric stealth, but robustness against fully adaptive attacks remains an open problem.

\section{Conclusion}
We presented a static, data-agnostic framework for detecting backdoors in LoRA adapters by analyzing the geometric and spectral structure of weight updates. By operating directly in weight space, the method avoids model execution and input data, enabling efficient pre-deployment screening at hub scale. Across held-out benchmarks for Llama, Qwen, and Gemma, the proposed detector achieves perfect separability between benign and poisoned adapters.

This work demonstrates that backdoor behaviours leave identifiable spectral signatures in parameter-efficient adaptations, and that weight-space analysis provides a principled and practical alternative to execution-based defences. More broadly, the experiments position attention-projection-wise geometric analysis of adapter weights as a promising direction for securing the emerging ecosystem of reusable PEFT components in large language models. Future work includes studying adaptive adversaries, testing stronger distribution shifts, and validating the detector on additional architectures and attack strategies.

\FloatBarrier
\clearpage

\bibliography{main}

@article{cox1958regression,
  author  = {Cox, D. R.},
  title   = {The Regression Analysis of Binary Sequences},
  journal = {Journal of the Royal Statistical Society. Series B (Methodological)},
  volume  = {20},
  number  = {2},
  pages   = {215--242},
  year    = {1958},
  url     = {https://www.jstor.org/stable/2983890}
}

@article{halko2011finding,
  title         = {Finding Structure with Randomness: Probabilistic Algorithms for Constructing Approximate Matrix Decompositions},
  author        = {Halko, Nathan and Martinsson, Per-Gunnar and Tropp, Joel A.},
  journal       = {SIAM Review},
  volume        = {53},
  number        = {2},
  pages         = {217--288},
  year          = {2011},
  doi           = {10.1137/090771806},
  eprint        = {0909.4061},
  archivePrefix = {arXiv},
  primaryClass  = {math.NA},
  url           = {https://arxiv.org/abs/0909.4061}
}

@article{llama3,
  title={The Llama 3 Herd of Models},
  author={{Llama Team}},
  journal={arXiv:2407.21783},
  url = {https://ai.meta.com/research/publications/the-llama-3-herd-of-models/},
  year={2024}
}

@article{qwen25,
  title         = {Qwen2.5 Technical Report},
  author        = {{Qwen Team}},
  journal       = {arXiv preprint arXiv:2412.15115},
  year          = {2024},
  eprint        = {2412.15115},
  archivePrefix = {arXiv},
  primaryClass  = {cs.CL},
  url           = {https://arxiv.org/abs/2412.15115}
}

@article{gemma2,
  title         = {Gemma 2: Improving Open Language Models at a Practical Size},
  author        = {{Gemma Team} and Mesnard, Thomas and Hardin, Cody and Dadashi, Robert and Bhupatiraju, Surya and Pathak, Shreya and Sifre, Laurent and Riv{\'e}re, Morgane and Kale, Mihir and Love, Julien and Tafti, Pooya and Hussenot, L{\'e}onard and Sessa, Pier Giuseppe and Chowdhery, Aakanksha and Welbl, Johannes and Pillai, TS Jayram and Manzini, Thomas and Besiroglu, Tamay and Cornebise, Julien and Konecny, Jakub and Basilico, Justin and Hoffmann, Matthew and Borgeaud, Sebastian and Meng, Yury and Hennigan, Tom and Elsen, Erich and Dean, Jeffrey and Kavukcuoglu, Koray and Hassabis, Demis and Barnard, Sam and Clark, Aidan and de Las Casas, Diego and Guy, Anjuli and Osindero, Simon and Simonyan, Karen and Vinyals, Oriol and Pascanu, Razvan and Viola, Fabio and Harvey, Neil and Johnson, Melvin and Bingham, George and Winter, Clemens and Sreevatsa, Sudarshan and Wang, Lily and many others},
  journal       = {arXiv preprint arXiv:2408.00118},
  year          = {2024},
  eprint        = {2408.00118},
  archivePrefix = {arXiv},
  primaryClass  = {cs.CL},
  url           = {https://arxiv.org/abs/2408.00118}
}

@article{youden_j,
    author = {Youden, W. J.},
    title = {Index for rating diagnostic tests},
    journal = {Cancer},
    volume = {3},
    number = {1},
    pages = {32--35},
    year = {1950},
    doi = {10.1002/1097-0142(1950)3:1<32::AID-CNCR2820030106>3.0.CO;2-3},
    url = {https://acsjournals.onlinelibrary.wiley.com/doi/10.1002/1097-0142(1950)3:1%3C32::AID-CNCR2820030106%3E3.0.CO;2-3}
}

@article{kurita2020weight,
  author       = {Keita Kurita and
                  Paul Michel and
                  Graham Neubig},
  title        = {Weight Poisoning Attacks on Pre-trained Models},
  journal      = {CoRR},
  volume       = {abs/2004.06660},
  year         = {2020},
  url          = {https://arxiv.org/abs/2004.06660},
  eprinttype    = {arXiv},
  eprint       = {2004.06660},
  bibsource    = {dblp computer science bibliography, https://dblp.org}
}

@article{hu2021lora,
        title={LoRA: Low-Rank Adaptation of Large Language Models}, 
      author={Edward J. Hu and Yelong Shen and Phillip Wallis and Zeyuan Allen-Zhu and Yuanzhi Li and Shean Wang and Lu Wang and Weizhu Chen},
      year={2021},
      eprint={2106.09685},
      archivePrefix={arXiv},
      primaryClass={cs.CL},
      url={https://arxiv.org/abs/2106.09685}, 
}

@article{tran2018spectral,
      title={Spectral Signatures in Backdoor Attacks}, 
      author={Brandon Tran and Jerry Li and Aleksander Madry},
      year={2018},
      eprint={1811.00636},
      archivePrefix={arXiv},
      primaryClass={cs.LG},
      url={https://arxiv.org/abs/1811.00636}, 
}

@article{gu2017badnets,
    title={BadNets: Identifying Vulnerabilities in the Machine Learning Model Supply Chain}, 
      author={Tianyu Gu and Brendan Dolan-Gavitt and Siddharth Garg},
      year={2019},
      eprint={1708.06733},
      archivePrefix={arXiv},
      primaryClass={cs.CR},
      url={https://arxiv.org/abs/1708.06733},
}

@article{chen2018detection,
        title={Detecting Backdoor Attacks on Deep Neural Networks by Activation Clustering}, 
      author={Bryant Chen and Wilka Carvalho and Nathalie Baracaldo and Heiko Ludwig and Benjamin Edwards and Taesung Lee and Ian Molloy and Biplav Srivastava},
      year={2018},
      eprint={1811.03728},
      archivePrefix={arXiv},
      primaryClass={cs.LG},
      url={https://arxiv.org/abs/1811.03728}, 
}

@article{luong2026lora,
      title={Why LoRA Fails to Forget: Regularized Low-Rank Adaptation Against Backdoors in Language Models}, 
      author={Hoang-Chau Luong and Lingwei Chen},
      year={2026},
      eprint={2601.06305},
      archivePrefix={arXiv},
      primaryClass={cs.CL},
      url={https://arxiv.org/abs/2601.06305}, 
}

@inproceedings{sun2024peftguard,
  title     = {PEFTGuard: Detecting Backdoor Attacks Against Parameter-Efficient Fine-Tuning},
  author    = {Sun, Zhen and Cong, Tianshuo and Liu, Yule and Lin, Chenhao and He, Xinlei and Chen, Rongmao and Han, Xingshuo and Huang, Xinyi},
  booktitle = {2025 IEEE Symposium on Security and Privacy (SP)},
  pages     = {1713--1731},
  year      = {2025},
  month     = may,
  publisher = {IEEE},
  doi       = {10.1109/SP61157.2025.00161},
  url       = {https://doi.org/10.1109/SP61157.2025.00161}
}

@article{auditing2025,
  title         = {A General Framework for Data-Use Auditing of {ML} Models},
  author        = {Huang, Zonghao and Gong, Neil Zhenqiang and Reiter, Michael K.},
  year          = {2025},
  eprint        = {2407.15100},
  archivePrefix = {arXiv},
  primaryClass  = {cs.CL},
  url           = {https://arxiv.org/abs/2407.15100}
}

@article{wang2025datafilter,
  title         = {Defending Against Prompt Injection with {DataFilter}},
  author        = {Wang, Yizhu and Chen, Sizhe and Alkhudair, Raghad and Alomair, Basel and Wagner, David},
  year          = {2025},
  eprint        = {2510.19207},
  archivePrefix = {arXiv},
  primaryClass  = {cs.CR},
  url           = {https://arxiv.org/abs/2510.19207}
}

@phdthesis{sperl2023activationanalysis,
  title         = {Defending Neural Networks with Activation Analysis},
  author        = {Sperl, Philip},
  school        = {Technische Universit{\"a}t M{\"u}nchen},
  year          = {2023},
  month         = {Apr},
  url           = {https://mediatum.ub.tum.de/doc/1700602/1700602.pdf}
}

@online{huggingface_hub_docs,
  author = {HF},
  title  = {Hugging Face Hub Documentation},
  year   = {2026},
  url    = {https://huggingface.co/docs/hub/index},
  note   = {Accessed: February 3, 2026},
}

@inproceedings{xu2024ltdefense,
  title     = {{LT-Defense}: Searching-free Backdoor Defense via Exploiting the Long-tailed Effect},
  author    = {Xu, Yixiao and Fang, Binxing and Li, Mohan and Tang, Keke and Tian, Zhihong},
  booktitle = {Advances in Neural Information Processing Systems},
  year      = {2024},
  url       = {https://papers.nips.cc/paper_files/paper/2024/file/064f6bcd7d3c72fb187bfca35ba2bfd4-Paper-Conference.pdf}
}

@article{chaudhary2025evaluation,
  title={Evaluation Awareness Scales Predictably in Open-Weights Large Language Models},
  author={Chaudhary, Maheep and Su, Ian and Hooda, Nikhil and Shankar, Nishith and Tan, Julia and Zhu, Kevin and Lagasse, Ryan and Sharma, Vasu and Panda, Ashwinee},
  journal={arXiv preprint arXiv:2509.13333},
  year={2025}
}

@article{balzano2025lowrank,
  title         = {An Overview of Low-Rank Structures in the Training and Adaptation of Large Models},
  author        = {Balzano, Laura and Ding, Tianjiao and Haeffele, Benjamin D. and Kwon, Soo Min and Qu, Qing and Wang, Peng and Wang, Zhangyang and Yaras, Can},
  journal       = {arXiv preprint arXiv:2503.19859},
  year          = {2025},
  eprint        = {2503.19859},
  archivePrefix = {arXiv},
  primaryClass  = {cs.LG},
  url           = {https://arxiv.org/abs/2503.19859}
}

@inproceedings{liu2025loratk,
  title     = {{LoRATK}: {LoRA} Once, Backdoor Everywhere in the Share-and-Play Ecosystem},
  author    = {Liu, Hongyi and Zhong, Shaochen and Sun, Xintong and Tian, Minghao and Hariri, Mohsen and Liu, Zirui and Tang, Ruixiang and Jiang, Zhimeng and Yuan, Jiayi and Chuang, Yu-Neng and Li, Li and Choi, Soo-Hyun and Chen, Rui and Chaudhary, Vipin and Hu, Xia},
  booktitle = {Findings of the Association for Computational Linguistics: EMNLP 2025},
  pages     = {23009--23047},
  year      = {2025},
  publisher = {Association for Computational Linguistics},
  doi       = {10.18653/v1/2025.findings-emnlp.1253},
  url       = {https://aclanthology.org/2025.findings-emnlp.1253/}
}

@article{wang2024modelsupply,
  title         = {Model Supply Chain Poisoning: Backdooring Pre-trained Models via Embedding Indistinguishability},
  author        = {Wang, Hao and Guo, Shangwei and He, Jialing and Liu, Hangcheng and Zhang, Tianwei and Xiang, Tao},
  journal       = {arXiv preprint arXiv:2401.15883},
  year          = {2024},
  eprint        = {2401.15883},
  archivePrefix = {arXiv},
  primaryClass  = {cs.CR},
  url           = {https://arxiv.org/abs/2401.15883}
}

@inproceedings{wang2019neural,
  title     = {Neural Cleanse: Identifying and Mitigating Backdoor Attacks in Neural Networks},
  author    = {Wang, Bolun and Yao, Yuanshun and Shan, Shawn and Li, Huiying and Viswanath, Bimal and Zheng, Haitao and Zhao, Ben Y.},
  booktitle = {2019 IEEE Symposium on Security and Privacy (SP)},
  year      = {2019},
  publisher = {IEEE},
  url       = {https://ieeexplore.ieee.org/document/8835365}
}

@inproceedings{gao2019strip,
  title     = {{STRIP}: A Defence Against Trojan Attacks on Deep Neural Networks},
  author    = {Gao, Yansong and Xu, Chang and Wang, Derui and Chen, Shiping and Ranasinghe, Damith C. and Nepal, Surya},
  booktitle = {Proceedings of the 35th Annual Computer Security Applications Conference},
  pages     = {113--125},
  year      = {2019},
  publisher = {Association for Computing Machinery},
  doi       = {10.1145/3359789.3359791},
  url       = {https://arxiv.org/abs/1902.06531}
}

@inproceedings{hayase2021spectre,
  title     = {{SPECTRE}: Defending Against Backdoor Attacks Using Robust Statistics},
  author    = {Hayase, Jonathan and Kong, Weihao and Somani, Raghav and Oh, Sewoong},
  booktitle = {Proceedings of the 38th International Conference on Machine Learning},
  series    = {Proceedings of Machine Learning Research},
  volume    = {139},
  pages     = {4129--4139},
  year      = {2021},
  publisher = {PMLR},
  url       = {https://proceedings.mlr.press/v139/hayase21a.html}
}

@article{qiu2022critical,
  title         = {Towards A Critical Evaluation of Robustness for Deep Learning Backdoor Countermeasures},
  author        = {Qiu, Huming and Ma, Hua and Zhang, Zhi and Abuadbba, Alsharif and Kang, Wei and Fu, Anmin and Gao, Yansong},
  journal       = {arXiv preprint arXiv:2204.06273},
  year          = {2022},
  eprint        = {2204.06273},
  archivePrefix = {arXiv},
  primaryClass  = {cs.CR},
  url           = {https://arxiv.org/abs/2204.06273}
}

@article{xu2025stealthiness,
  title         = {Towards Backdoor Stealthiness in Model Parameter Space},
  author        = {Xu, Xiaoyun and Liu, Zhuoran and Koffas, Stefanos and Picek, Stjepan},
  journal       = {arXiv preprint arXiv:2501.05928},
  year          = {2025},
  eprint        = {2501.05928},
  archivePrefix = {arXiv},
  primaryClass  = {cs.CR},
  url           = {https://arxiv.org/abs/2501.05928}
}

@misc{taori2023alpaca,
  title        = {Alpaca: A Strong, Replicable Instruction-Following Model},
  author       = {Taori, Rohan and Gulrajani, Ishaan and Zhang, Tianyi and Dubois, Yann and Li, Xuechen and Guestrin, Carlos and Liang, Percy and Hashimoto, Tatsunori B.},
  year         = {2023},
  howpublished = {Stanford Center for Research on Foundation Models},
  url          = {https://crfm.stanford.edu/2023/03/13/alpaca}
}

@misc{databricks2023dolly15k,
  title        = {Databricks Dolly 15k},
  author       = {{Databricks}},
  year         = {2023},
  howpublished = {Hugging Face dataset card},
  url          = {https://huggingface.co/datasets/databricks/databricks-dolly-15k}
}

@article{cobbe2021gsm8k,
  title         = {Training Verifiers to Solve Math Word Problems},
  author        = {Cobbe, Karl and Kosaraju, Vineet and Bavarian, Mohammad and Chen, Mark and Jun, Heewoo and Kaiser, {\L}ukasz and Plappert, Matthias and Tworek, Jerry and Hilton, Jacob and Nakano, Reiichiro and Hesse, Christopher and Schulman, John},
  journal       = {arXiv preprint arXiv:2110.14168},
  year          = {2021},
  eprint        = {2110.14168},
  archivePrefix = {arXiv},
  primaryClass  = {cs.LG},
  url           = {https://arxiv.org/abs/2110.14168}
}

@article{clark2018arc,
  title         = {Think You Have Solved Question Answering? Try {ARC}, the {AI2} Reasoning Challenge},
  author        = {Clark, Peter and Cowhey, Isaac and Etzioni, Oren and Khot, Tushar and Sabharwal, Ashish and Schoenick, Carissa and Tafjord, Oyvind and Turney, Peter and Khashabi, Daniel},
  journal       = {arXiv preprint arXiv:1803.05457},
  year          = {2018},
  eprint        = {1803.05457},
  archivePrefix = {arXiv},
  primaryClass  = {cs.AI},
  url           = {https://arxiv.org/abs/1803.05457}
}

@inproceedings{rajpurkar2018squad2,
  title     = {Know What You Don't Know: Unanswerable Questions for {SQuAD}},
  author    = {Rajpurkar, Pranav and Jia, Robin and Liang, Percy},
  booktitle = {Proceedings of the 56th Annual Meeting of the Association for Computational Linguistics},
  volume    = {2},
  pages     = {784--789},
  year      = {2018},
  publisher = {Association for Computational Linguistics},
  doi       = {10.18653/v1/P18-2124},
  url       = {https://aclanthology.org/P18-2124/}
}

@article{kwiatkowski2019naturalquestions,
  title     = {Natural Questions: A Benchmark for Question Answering Research},
  author    = {Kwiatkowski, Tom and Palomaki, Jennimaria and Redfield, Olivia and Collins, Michael and Parikh, Ankur and Alberti, Chris and Epstein, Danielle and Polosukhin, Illia and Devlin, Jacob and Lee, Kenton and Toutanova, Kristina and Jones, Llion and Kelcey, Matthew and Chang, Ming-Wei and Dai, Andrew M. and Uszkoreit, Jakob and Le, Quoc and Petrov, Slav},
  journal   = {Transactions of the Association for Computational Linguistics},
  volume    = {7},
  pages     = {453--466},
  year      = {2019},
  publisher = {MIT Press},
  doi       = {10.1162/tacl_a_00276},
  url       = {https://aclanthology.org/Q19-1026/}
}

@article{chen2021humaneval,
  title         = {Evaluating Large Language Models Trained on Code},
  author        = {Chen, Mark and Tworek, Jerry and Jun, Heewoo and Yuan, Qiming and Pinto, Henrique Ponde de Oliveira and Kaplan, Jared and Edwards, Harri and Burda, Yuri and Joseph, Nicholas and Brockman, Greg and Ray, Alex and Puri, Raul and Krueger, Gretchen and Petrov, Michael and Khlaaf, Heidy and Sastry, Girish and Mishkin, Pamela and Chan, Brooke and Gray, Scott and Ryder, Nick and Pavlov, Mikhail and Power, Alethea and Kaiser, {\L}ukasz and Bavarian, Mohammad and Winter, Clemens and Tillet, Philippe and Such, Felipe Petroski and Cummings, Dave and Plappert, Matthias and Chantzis, Fotios and Barnes, Elizabeth and Herbert-Voss, Ariel and Guss, William and Nichol, Alex and Paino, Alex and Tezak, Nikolas and Tang, Jie and Babuschkin, Igor and Balaji, Suchir and Jain, Shantanu and Saunders, William and Hesse, Christopher and Carr, Andrew N. and Leike, Jan and Achiam, Josh and Misra, Vedant and Morikawa, Evan and Radford, Alec and Knight, Matthew and Brundage, Miles and Murati, Mira and Mayer, Katie and Welinder, Peter and McGrew, Bob and Amodei, Dario and McCandlish, Sam and Sutskever, Ilya and Zaremba, Wojciech},
  journal       = {arXiv preprint arXiv:2107.03374},
  year          = {2021},
  eprint        = {2107.03374},
  archivePrefix = {arXiv},
  primaryClass  = {cs.LG},
  url           = {https://arxiv.org/abs/2107.03374}
}

@inproceedings{wang2018glue,
  title     = {{GLUE}: A Multi-Task Benchmark and Analysis Platform for Natural Language Understanding},
  author    = {Wang, Alex and Singh, Amanpreet and Michael, Julian and Hill, Felix and Levy, Omer and Bowman, Samuel R.},
  booktitle = {Proceedings of the 2018 {EMNLP} Workshop {BlackboxNLP}: Analyzing and Interpreting Neural Networks for {NLP}},
  pages     = {353--355},
  year      = {2018},
  publisher = {Association for Computational Linguistics},
  doi       = {10.18653/v1/W18-5446},
  url       = {https://aclanthology.org/W18-5446/}
}
\bibliographystyle{colm2026_conference}

\appendix

\section{Appendix}\label{sec:appendix_experiments}

\subsection{Calibration and Held-Out Score Distributions}
Figure~\ref{fig:appendix_calibration_distribution} reports the calibration score distribution on the validation subset of the calibration set.
In all three model families, the selected threshold lies inside a visible separation margin, which is consistent with the strict separation later observed on the held-out test split.

\begin{figure*}[t]
    \centering
    \includegraphics[width=0.96\textwidth]{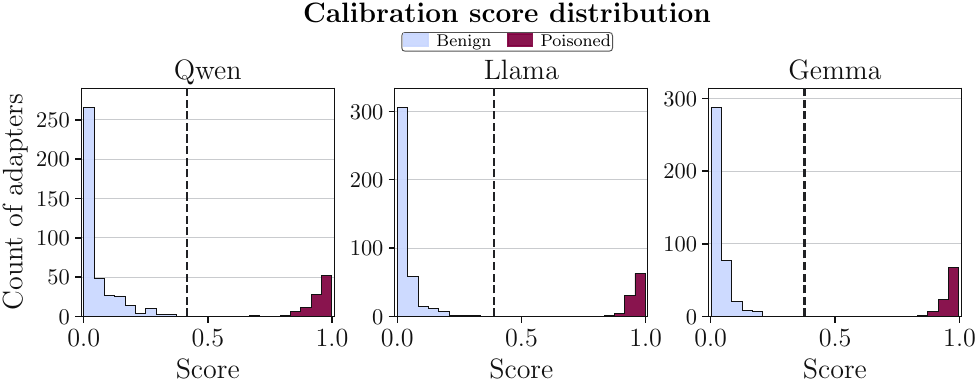}
    \caption{\textbf{Calibration score distribution across the three model families.}
    All three backbones exhibit a clear validation-time gap between benign and poisoned scores before the detector is evaluated.}
    \label{fig:appendix_calibration_distribution}
\end{figure*}

\subsection{Layer and Rank Sweep}\label{sec:appendix_layer_rank}
Section~\ref{sec:robustness_layer_rank} uses Figure~\ref{fig:layer_rank_sensitivity} to state the central Q3 conclusion: the detection separability is more sensitive to layer placement than to moderate changes in rank.
This subsection reports the raw probe-based operating-point diagnostics and the per-metric layer/rank heatmaps underlying that summary.
Taken together, the appendix figures support two distinct points:
\begin{enumerate}
    \item First, they show that the raw probe ROC-AUC, transfer estimate, symmetric KL, and spectral \(|\Delta W|\) vary across layer and rank before those quantities are aggregated into the composite score.
    \item Second, they show that the preference for late layers is supported jointly by probe ROC-AUC, symmetric KL, and spectral \(|\Delta W|\), rather than by any one of these quantities alone.
\end{enumerate}

The composite score summarized in Figure~\ref{fig:layer_rank_sensitivity} and discussed in Section~\ref{sec:robustness_layer_rank} is defined as
\begin{equation}
C(\ell,r)=\frac{1}{|\mathcal{M}|}\sum_{m\in\mathcal{M}}\operatorname{norm}\!\big(m(\ell,r)\big),
\end{equation}
\begin{equation}
    \mathcal{M}=\{\text{ROC-AUC},\text{Probe Acc},\text{Symmetric KL},\text{Spectral }|\Delta W|\},
\end{equation}
where \(\operatorname{norm}(\cdot)\) denotes min--max normalization over the tested layer/rank grid.
This score is used only to summarize the operating-point preference in Figure~\ref{fig:layer_rank_sensitivity}; the appendix figures below expose the same sweep before aggregation.

\paragraph{Experiment 1: Raw operating-point diagnostics.}
Figure~\ref{fig:appendix_lora_placement} supports the first point above and reports the raw probe-based operating-point diagnostics.
Panel~(a) measures local probe ROC-AUC at the finetuned layer and shows a following effect: along the \(r=16\) sweep, later layers are consistently more separable than early layers, while the rank sweep at layer 21 produces only modest variation.
Panel~(b) reports an auxiliary off-layer transfer proxy computed from non-finetuned layers.
Because this proxy does not track local separability monotonically, we treat it as mechanistic context, not the selection criterion for the benchmark operating point.

\paragraph{Experiment 2: Per-metric decomposition of the sweep.}
Figure~\ref{fig:appendix_layer_rank_full} supports the second point above by decomposing the composite layer/rank heatmap from Figure~\ref{fig:layer_rank_sensitivity} into its constituent metrics.
The same qualitative pattern remains visible across probe ROC-AUC, probe accuracy, symmetric KL, and spectral \( |\Delta W| \): late-layer placements are systematically stronger than early-layer placements, whereas the tested rank changes remain secondary.
Individual metrics still favor slightly different cells inside the late-layer regime, but the aggregate picture is consistent with the operating-point choice stated in Section~\ref{sec:robustness_layer_rank}: layer 21 with rank 16.

\subsection{Projection-Wise Metric Panels}\label{sec:appendix_projection_panels}
Figures~\ref{fig:appendix_sigma1_panels}--\ref{fig:appendix_kurtosis_panels} show the full projection-wise feature distributions for each metric.
Each figure stacks Qwen, Llama, and Gemma vertically and compares the four LoRA projections within each model family, together with the one-dimensional ROC-AUC of the corresponding $(\text{projection},\text{metric})$ pair.
These plots support the two takeaways stated in Section~\ref{sec:feature_projection_structure}:
single cues can be highly informative, but the strongest cue varies by backbone; and no single scalar feature is uniformly dominant across all three architectures.

\begin{figure}[p]
    \centering
    \includegraphics[width=\linewidth]{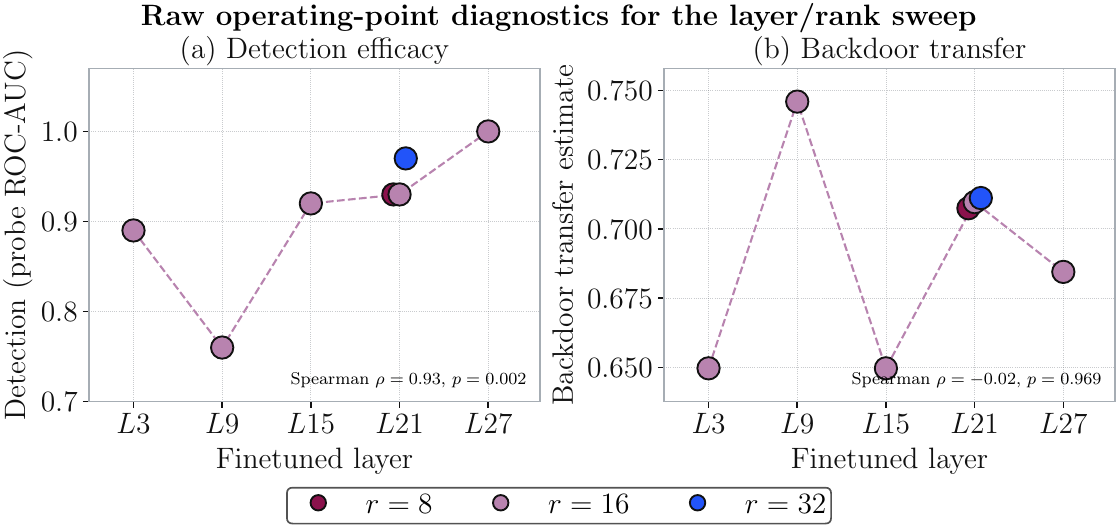}
    \caption{\textbf{Raw operating-point diagnostics for the layer/rank sweep.}
    \textbf{(a)} Local probe ROC-AUC at the finetuned layer.
    \textbf{(b)} Auxiliary off-layer transfer proxy computed from non-finetuned layers.
    The two panels support the same main interpretation: local separability improves substantially with depth, whereas rank has a weaker effect over the tested range.}
    \label{fig:appendix_lora_placement}
\end{figure}

\begin{figure*}[p]
    \centering
    \includegraphics[width=0.95\textwidth]{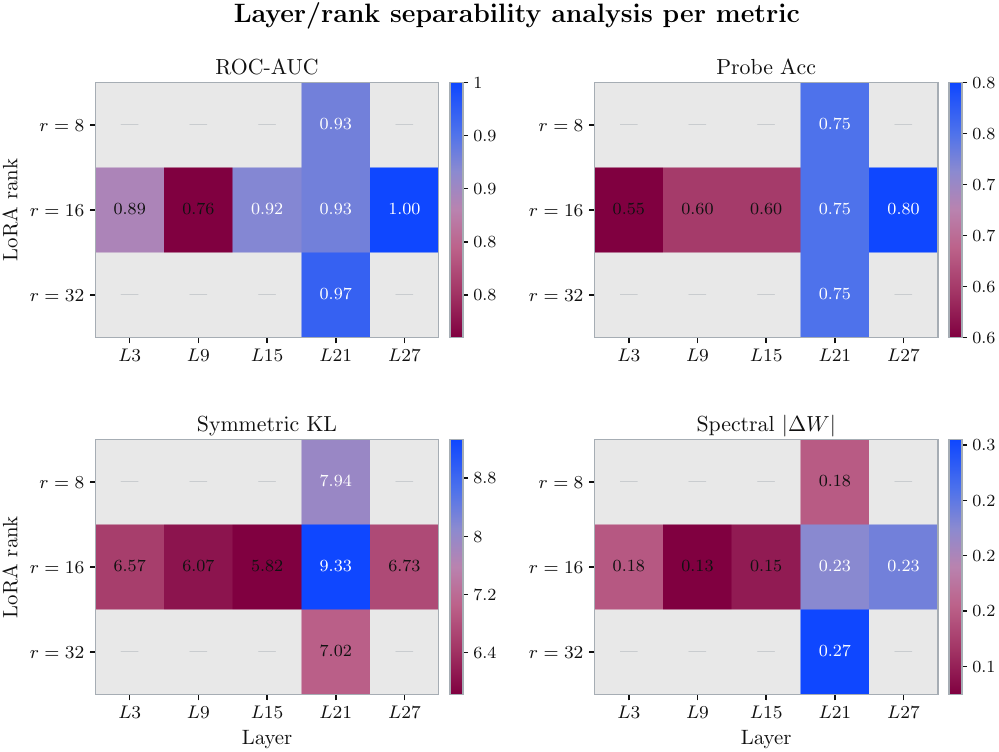}
    \caption{\textbf{Per-metric decomposition of the layer/rank sweep.}
    Grey cells denote untested configurations.
    Each panel shows one separability criterion over the same grid used in Figure~\ref{fig:layer_rank_sensitivity}.
    The late-layer preference is visible across all four metrics, although the strongest individual cell varies slightly by metric.}
    \label{fig:appendix_layer_rank_full}
\end{figure*}

\begin{figure*}[p]
    \centering
    \includegraphics[width=0.96\textwidth]{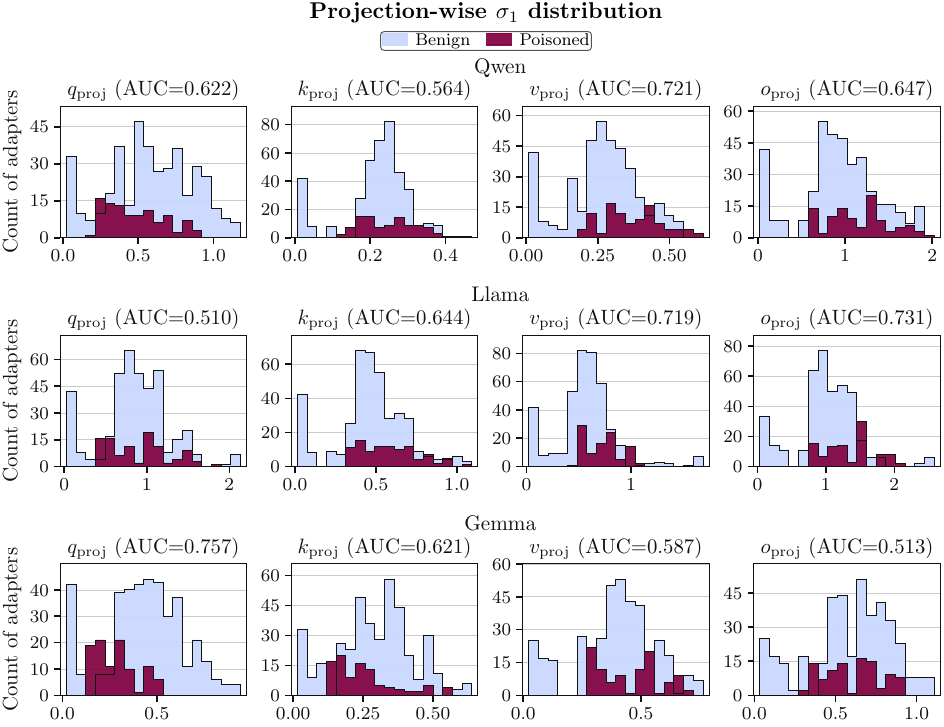}
    \caption{\textbf{Projection-wise $\sigma_1$ distribution across model families.}
    For each backbone, the four columns correspond to the $q$, $k$, $v$, and $o$ projections.}
    \label{fig:appendix_sigma1_panels}
\end{figure*}

\begin{figure*}[p]
    \centering
    \includegraphics[width=0.96\textwidth]{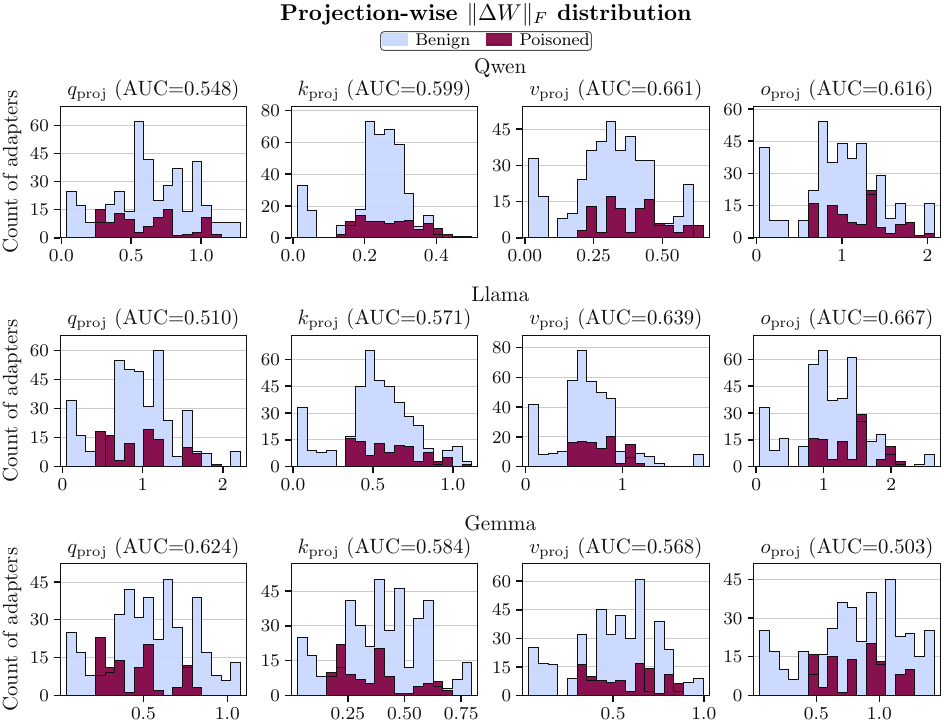}
    \caption{\textbf{Projection-wise Frobenius-norm distribution across model families.}
    Rows correspond to model families; columns correspond to the four LoRA projections.}
    \label{fig:appendix_frobenius_panels}
\end{figure*}

\begin{figure*}[p]
    \centering
    \includegraphics[width=0.94\textwidth]{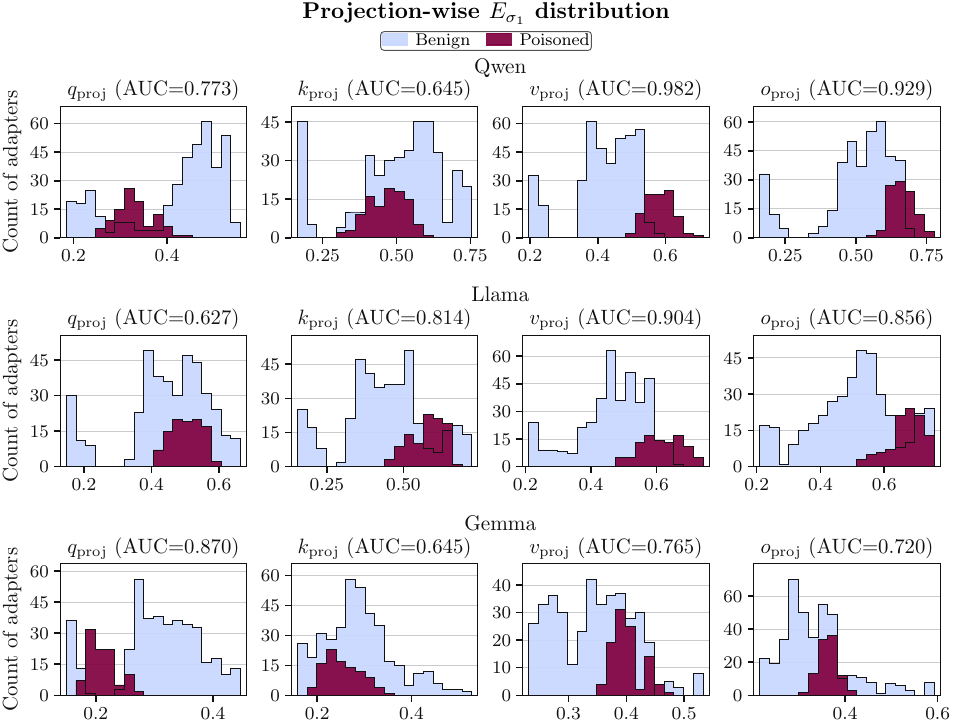}
    \caption{\textbf{Projection-wise energy-concentration distribution across model families.}}
    \label{fig:appendix_energy_panels}
\end{figure*}

\begin{figure*}[p]
    \centering
    \includegraphics[width=0.94\textwidth]{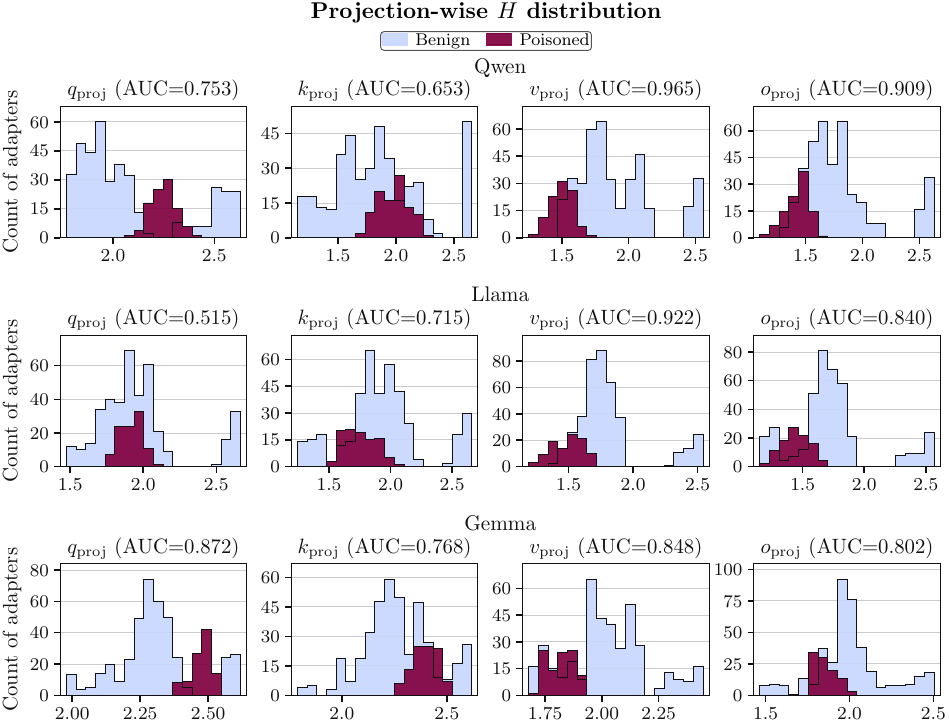}
    \caption{\textbf{Projection-wise entropy distribution across model families.}}
    \label{fig:appendix_entropy_panels}
\end{figure*}

\clearpage
\noindent
\begin{minipage}{\textwidth}
    \centering
    \includegraphics[width=0.96\textwidth]{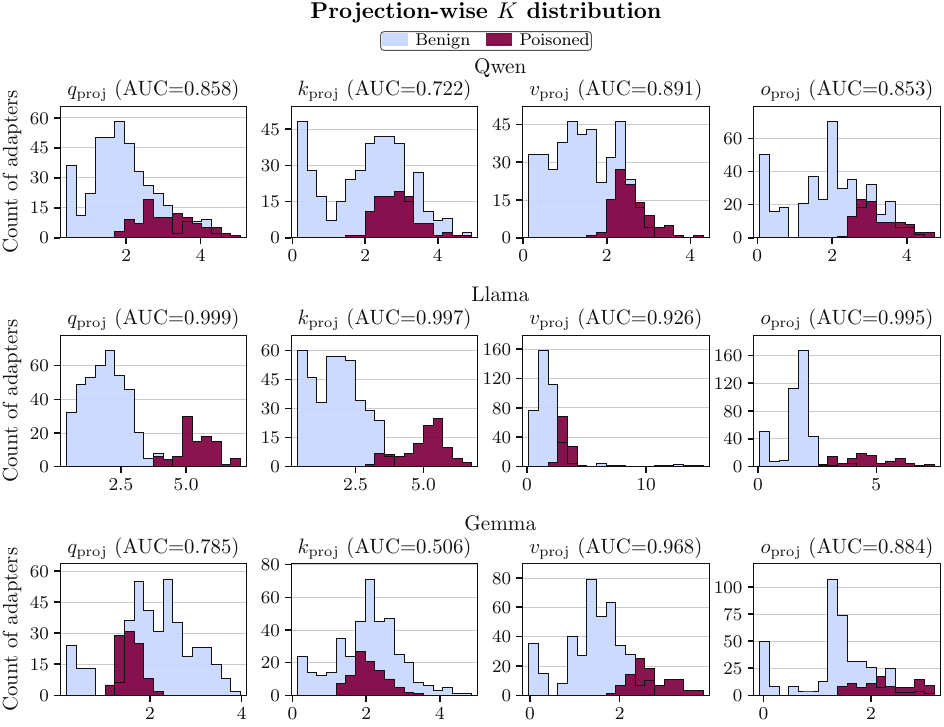}
    \captionof{figure}{\textbf{Projection-wise kurtosis distribution across model families.}
    Strong separation in some rows highlights that informative one-dimensional cues are architecture-dependent.}
    \label{fig:appendix_kurtosis_panels}
\end{minipage}

\end{document}